\documentstyle[aps,pra,subfigure,epsfig,amsmath,amssymb]{revtex}

\begin{document}
\draft

\title{Dynamic structure factor of  a superfluid Fermi gas}
\author{Anna Minguzzi$^{*}$, Gabriele Ferrari$^{\ddagger,\S}$
  and Yvan Castin$^{\ddagger}$}

\address{$^*$ Istituto Nazionale di Fisica della Materia and
Classe di Scienze, Scuola Normale Superiore, \\ Piazza dei Cavalieri 7,
I-56126 Pisa, Italy \\
$^\S$ Istituto Nazionale di Fisica della Materia and European
Laboratory for Non-Linear Spectroscopy, Universit\`a di Firenze, Largo E. Fermi, 2, I-50125
Firenze, Italy  \\
$^\ddagger$ Laboratoire Kastler-Brossel, Ecole Normale Sup\'erieure, 24 rue
Lhomond, F-75231 Paris, France\\
}

\maketitle                
\begin{abstract}
We describe the excitation spectrum of a two-component neutral Fermi
gas in the superfluid phase at finite temperature by deriving a
suitable Random-Phase approximation with the technique of functional
derivatives. The obtained spectrum for the homogeneous gas at small
wavevectors contains the Bogoliubov-Anderson phonon and is essentially
different from the spectrum predicted by the static Bogoliubov theory,
which instead shows an unphysically
large response. We adapt the results for the
homogeneous system to obtain the  dynamic structure factor
of a harmonically confined superfluid 
and we identify in the spectrum 
 a unique feature of the
superfluid phase.
\end{abstract}


\section{Introduction}
The techniques of atom trapping and cooling which have led to the
realization of Bose-Einstein condensation in alkali gases are
currently being employed to cool also the fermionic isotopes $^{40}$K
\cite{fermi_degen} 
and $^6$Li \cite{gabriele}.  Dilute gases of fermionic atoms with attractive
inter-particle interactions are predicted to undergo a superfluid
transition at low temperatures ($T_{sup} \ll T_F$, where $T_F$ is the Fermi
temperature). The simplest mechanism envisaged is a $s$-wave pairing
\cite{stoof_prl,stoof_marianne} 
which can be obtained, compatibly with the Pauli principle, between
atoms belonging to two different internal states. 
The realization of a two-component Fermi gas in the superfluid state
may provide a new physical system to study. Its  properties are
expected to be different from those of superfluid $^3$He,
which has a $p$-wave pairing and is not in the dilute regime, and from
conventional charged superconductors which have an excitation
spectrum dominated by the Coulomb interaction \cite{martin_didactic}
 and only weakly modified by the superfluid transition. 

An important issue for future experiments is to identify a clear
signature of the superfluid transition in an atomic Fermi gas. Contrary to
the case of atomic Bose-Einstein condensates, for fermions  
the superfluid transition affects only slightly the density
profile and the internal energy of the gas \cite{commento}.
A first idea is to measure the pair distribution function of the
atoms, {\it e.g.} by using a laser probe beam \cite{bcs_1}.
A second idea is to look at the dynamical properties,
which are expected to be dramatically modified by the transition.
Several proposals have been put forward  in this direction, such as 
Cooper-pair
breaking {\it via} a Raman transition \cite{bcs_2}, measurement 
of the moment of inertia of the cloud \cite{schuck}, and
 excitation of 
 collective modes in a harmonic trap 
by modulation of the
trap frequencies \cite{baranov} or rotation of the axis of the trap
\cite{anna}.

In this work we suggest to identify the superfluid phase
through the measurement  of the 
bulk excitations of the gas, {\it i.e.}
excitations with a wavelength smaller than the spatial extension of
the atomic cloud. This is complementary to the proposals in
\cite{baranov,anna} as it deals with high energy excitations in a
quasi-homogeneous system at arbitrary temperature, and is not
irrealistic from the experimental point of view, since
efficient Bragg scattering techniques have already been  successfully used 
to measure the excitation spectra of Bose
condensates \cite{ketterle}.

We obtain the
excitation spectrum of the fluid in the dilute regime by employing 
the Random-Phase (or Time-Dependent
Hartree-Fock-Gorkov) Approximation (RPA), which we derive explicitly
for the two-component system of present interest by the technique of
functional derivatives \cite{kadanoff_baym,griffin_rpa}.
We do not use here the usual static Bogolubov approximation
\cite{vecchio} for two main reasons: ({\it i}) physically its
 excitation spectrum has a gap and
therefore ignores the branch of phonon-like excitations
(Bogolubov-Anderson phonon, \cite{bogolubov_sound,anderson_rpa,leggett})
expected on 
very general grounds to
show up in  homogeneous neutral superconductors with short range
interactions  \cite{martin_didactic}, and ({\it ii}) the
density-density response
function obtained within this approximation shows an unphysically
large response at small 
wavevectors and 
fails to satisfy the
$f$-sum rule, which is a requirement deriving from the local 
particle conservation law. 
 
 We find that the RPA spectrum 
of a homogeneous two-component Fermi gas in the
superfluid phase possesses the continuum of particle-hole
excitations and a peak corresponding to the
Bogoliubov-Anderson phonon;
 it satisfies the $f$-sum rule and includes naturally the
Landau damping of the phonon due to the interplay with thermal
excitations. 
We adapt the results obtained 
for the homogeneous system 
to describe harmonically confined gases by means of a local-density
approximation, which predicts a broadening of the spectrum by taking
into account the inhomogeneity of the density profile. We predict that the
Bogoliubov-Anderson phonon, which is
the main feature of the spectrum in the superfluid phase, would remain
visible even in the trapped cloud. 

The structure of this paper is as follows. In Sec. II we derive the density
response function of the fluid in the RPA.
In Sec. III we obtain the spectrum of density fluctuations first in the
homogeneous case and then for an  harmonically trapped gas. Finally,
Sec. IV gives a summary of our results and offers some concluding remarks.

\section{Random-Phase Approximation} 

We describe a two-component  atomic Fermi gas  by the
following Hamiltonian:
\begin{eqnarray}
\hat {\cal H} -\mu \hat {\cal N}&=&\sum_{\alpha=\uparrow,\downarrow}
\int d^3r\,  \hat
\psi^{\dag}_{\alpha}({\mathbf r})\left( -\frac{\hbar^2}{2m}\nabla
^2+V_{\mbox{\scriptsize ext},\alpha}({\mathbf r})+U({\mathbf r},t)
-\mu _{\alpha} \right)\hat \psi^{}_{\alpha}({\mathbf r}) \nonumber \\ &+&
\frac{1}{2} \sum_{\{\alpha,\beta\}=\uparrow,\downarrow}\int d^3r\int
d^3{r'}\,\hat\psi^{\dag}_{\alpha}({\mathbf r})\hat\psi^{\dag}_{\beta}({\mathbf r}')
v_{\alpha\beta}({\mathbf r}-{\mathbf r'})\hat\psi^{}_{\beta}({\mathbf
r}')\hat\psi^{}_{\alpha}({\mathbf r})  \;.
\end{eqnarray}
The fermionic field operators $\hat\psi^{}_{\alpha}({\mathbf r})$ satisfy the
usual anticommutation relations; the interactions are considered only
in $s$-wave between  fermions in different internal
states and are modeled by the  inter-particle potential
$v_{\alpha\beta}({\mathbf r}_1-{\mathbf 
r}_2)=\left(\begin{array}{cc}0 &
v_{\uparrow\downarrow}({\mathbf r}_1-{\mathbf
r}_2)\\v_{\uparrow\downarrow}({\mathbf 
r}_1-{\mathbf r}_2) & 0 \\
\end{array}\right)$, with $v_{\uparrow\downarrow}({\mathbf r}_1-{\mathbf
r}_2)=v_{\uparrow\downarrow} \delta({\mathbf r}_1-{\mathbf
 r}_2)\partial_{r_{12}}(r_{12}\cdot)$. This model interaction potential,
 known as the Fermi pseudopotential,
 leads to a divergence-free BCS theory \cite{yvan}.
For the sake of
generality we have included the  presence of the
external confinement  via  the  trapping potentials
$V_{\mbox{\scriptsize ext},\alpha}({\mathbf r})$ and the possibility of having
 different numbers 
of atoms in the two 
components; however in the following we shall restrict to the
derivation of the equations in the homogeneous system and in the
symmetric case $N_{\uparrow}=N_{\downarrow}=N/2$, which is the most
favourable for the formation of Cooper pairs \cite{stoof_marianne}.
The effect of trapping potential present in a realistic
experiment will be included later on with a local density approximation.
The external perturbing 
field  which is necessary to generate the total
density response  has 
been introduced as $U({\mathbf r},t)$.
Physically $U$ represents the action of the
probe applied in a real experiment; it may be a time-dependent
perturbation applied to the magnetic trap \cite{eccitazioni}, a probe
laser beam 
\cite{ketterle} or a test particle \cite{ketterle_test}.
Here we have assumed that the same potential $U$
acts  in the same way on both
components, and we shall determine the perturbation
$\delta\rho({\mathbf r},t)$
on the total density induced by the probe potential $U({\mathbf r},t)$,
assuming that the gas is initially at thermal equilibrium with a temperature 
$T$.   
We restrict to the linear response regime, where the density
perturbation is a linear functional of the probe potential
$U$ expressed through the density-density response function
$\chi({\mathbf r}_1, t_1, {\mathbf r}_2, t_2)$, a function of
two position vectors ${\mathbf r}_{1,2}$ and of two time variables
$t_{1,2}$:
\begin{equation}
\delta\rho({\mathbf r}_1,t_1) =  \int d^3r_2\, dt_2\
\chi({\mathbf r}_1, t_1, {\mathbf r}_2, t_2) U({\mathbf r}_2,t_2) .
\end{equation}
In this section we explain how to calculate this
response function in the Hartree-Fock-Gorkov approximation. We obtain
general equations valid for an arbitrary trapping potential 
$V_{\mbox{\scriptsize ext}}({\mathbf r})$. 
We then solve these equations explicitly
for a spatially homogeneous gas at thermal equilibrium, where $\chi$
is a function of ${\mathbf r}_1-{\mathbf r}_2$ and $t_1-t_2$
only.

To proceed with the derivation of the response function $\chi$,
we follow the imaginary time Green's function technique
of the book of Kadanoff and Baym \cite{kadanoff_baym}.
One first defines the two by two matrix of (normal and anomalous) Green's
functions in imaginary times
\begin{equation}
G(1,2)= \left(\begin{array}{cc}
G_{\uparrow\uparrow}(1,2) & G_{\uparrow\downarrow}(1,2) \\
G_{\downarrow\uparrow}(1,2)& G_{\downarrow\downarrow}(1,2) 
\end{array}\right)
\equiv -\langle {\rm T} \hat \Psi (1)
\hat \Psi ^{\dag}(2) \rangle
\end {equation}
where T is the time-ordering operator,
$\hat \Psi(1)=\left(\begin{array}{c} \hat\psi_{\uparrow}(1) \\
\hat\psi_{\downarrow}^{\dag}(1)\\
\end{array}\right)$,
$\hat \Psi^\dagger(2)=\left( \hat\psi_{\uparrow}^{\dag}(2),
\hat\psi_{\downarrow}(2)\right)$,
$\langle ...\rangle$ indicates the average over
the state of the system in the presence of the perturbing field $U$ 
and $(1,2)$ stands for $({\mathbf r}_1, i\tau_1,{\mathbf r}_2,
i\tau_2)$, where $\tau_1$ and $\tau_2$ are real quantities.
For  more details on the imaginary time technique,  we refer to
reference \cite{kadanoff_baym}.  We simply
note that the various functions considered here
can be obtained for real times by analytic continuation
of their imaginary time values.
 From the equation of motion for the field operator
in imaginary times 
 one derives \cite{kadanoff_baym} the generalized Dyson
equation for $ G$:
\begin{equation}
 G(1,2)= G_0(1,2)+ \int d\bar 3\int d\bar 4 \, G_0(1,\bar
3) \Sigma(\bar 3, \bar 4)  G(\bar 4,2)+ \int d\bar 3\, G_0(1,\bar
3) W(\bar 3)  G(\bar 3,2)\;,
\label{dyson}
\end{equation}
where the $2\times 2$ matrix
$ G_0(1,2)$ is the solution of the equations of motion in
absence of the interactions, $ W(1)
=\left(\begin{array}{cc} U(1) & 0 \\ 0 & -U(1)\\
\end{array}\right)$ is the $2\times 2$  matrix of external field 
and $ \Sigma(1,2)$ is the  $2\times 2$  matrix  of
self-energies.  Since we want to describe a dilute system, we work in 
the mean-field Hartree-Fock-Gorkov symmetry breaking
approximation, where the self-energy reads  
\begin{equation}
\Sigma_{HG}(1,2)=\delta(1,2) v_{\uparrow\downarrow}
\left(\begin{array}{cc}
\langle\hat{\psi}_{\downarrow}^\dagger(1)\hat{\psi}_{\downarrow}(1)\rangle
& 0 \\
0 &
-\langle\hat{\psi}_{\uparrow}^\dagger(1)\hat{\psi}_{\uparrow}(1)\rangle
\end{array} \right)
+v_{\uparrow\downarrow}(1,2)
\left(\begin{array}{cc}0 &G_{\uparrow\downarrow}(1,2) \\
G_{\downarrow\uparrow}(1,2) &0 
\end{array} \right).
\label{sigma_hg}
\end{equation}
Here we have used the notations
$\delta(1,2)= \delta(\vec{r}_1-\vec{r}_2)\delta(\tau_1-\tau_2)/i$
and $v_{\uparrow\downarrow}(1,2)=
v_{\uparrow\downarrow}({\mathbf r}_1,{\mathbf r}_2) \delta(\tau_1-\tau_2)/i$.
We remark that  the Fock contribution
is zero since the interaction takes place only between particles with
opposite spins and $\langle \psi^\dagger_\uparrow \psi_\downarrow\rangle=0$
in the considered state of the system, hence the vanishing diagonal in
the last term of  Eq.~(\ref{sigma_hg}).  

Following a standard approach \cite{kadanoff_baym,griffin_rpa}, we then
obtain the density response matrix in RPA by taking the
functional derivative of the Green's function with respect to the
external field $U$: we define the generalized response matrix as
the two by two matrix $ L(1,2,3)= \sigma_3 \delta  G(1,2) /\delta U(3)$,
where
$\sigma_3$ is the third Pauli matrix $\left(\begin{array}{cc}1 &0 \\ 
0 &-1 \\ \end{array} \right)$. 
The matrix giving the physical response is obtained from  
the limit $ L(1,2)\equiv L (1,1^+,2)$.
The density-density response function $\chi$ is simply the trace over the two
spin components of the response matrix:
\begin{equation}
\chi(1,2) = \mbox{Tr}\, L(1,2) = L_{\uparrow\uparrow}(1,2)
+L_{\downarrow\downarrow}(1,2).
\end{equation}

The equation for the density response in the Random-Phase
Approximation is obtained by the functional derivative of the Dyson
equation, Eq.~(\ref{dyson}), where the approximation (\ref{sigma_hg}) for the
self energy has been employed. This yields:
\begin{equation}
L(1,2)=L^0(1,2) + \frac{v_{\uparrow\downarrow}}{2} \int
d\bar 3\,  L^0(1,\bar 3) \chi(\bar 3,2)
-\int d\bar3\int d\bar4 \, \tilde G(1,\bar 3)
M(\bar 3,\bar 4,2) \tilde G(\bar 4,1)
v_{\uparrow\downarrow}(\bar 3,\bar4)
 \;.
\label{rpa_base}
\end{equation}
Here we have introduced the $2\times 2$ matrices
$\tilde G(1,2)=\sigma_3 G(1,2) $, $L^0(1,2)=\tilde G(1,2) \tilde G(2,1)$,
and $M(1,2,3)=\left(\begin{array}{cc}0 & 
L_{\uparrow\downarrow} (1,2,3) \\ 
L_{\downarrow\uparrow} (1,2,3) & 0\end{array}\right)$.
Physically $L^0(1,2)$ is the response matrix of the gas
in the static Bogoliubov approximation,
so that the reference system of the RPA
is not the ideal gas but the Bogoliubov gas of
quasiparticles. 

It is possible to display the diagrammatic structure of
Eq.~(\ref{rpa_base}) by separating out the ``proper'' part 
$\bar L(1,2)$ of the
density response. We have therefore an equation
which sums the bubble diagrams,
\begin{equation}   
L(1,2)= \bar L(1,2)+ \frac{1}{2} v_{\uparrow\downarrow}\int
d\bar3\, \bar L(1,\bar 3)\chi(\bar 3,2) \;,
\label{rpa_proper} 
\end{equation}
and an equation which defines the bubble as a sum of all the ladder
diagrams,
\begin{equation}
\bar L(1,2) =L^0(1,2)-\int
d\bar3\int d\bar4 \,\tilde G(1,\bar 3) v_{\uparrow\downarrow}(\bar 3,\bar 4) 
\bar M(\bar 3,\bar 4,2) \tilde G(\bar 4,1) 
\;.
\label{bolla}
\end{equation}
Here $\bar M=\left(\begin{array}{cc} 0 & \bar
L_{\uparrow\downarrow} \\ \bar
L_{\downarrow\uparrow} & 0 \end{array} \right)$.

We now specialize the previous equations to the case of a spatially
homogeneous gas 
and to the dilute limit $\Delta\ll \epsilon_F$, where $\Delta$ is the gap
and $\epsilon_F$ is the Fermi energy. All the response matrices 
depend then only on the relative spatial coordinates ${\mathbf r}=
{\mathbf r}_1-{\mathbf r}_2$
and the relative time coordinate $t=t_1-t_2$ and we introduce their
double Fourier transforms  
with respect to $\mathbf r$ and $t$, {\sl e.g.} 
\begin{equation}
\bar{L}({\mathbf q},\omega) = \int d^3r\, dt\, \bar{L}({\mathbf r}_1=
{\mathbf r},t_1=t,
{\mathbf r}_2={\mathbf 0},t_2=0) \, e^{i({\mathbf q}\cdot{\mathbf
r}-\omega t)}. 
\end{equation}
The following equation 
is obtained  for $\bar{L}({\mathbf q},\omega)$  from the
solution of Eq.~(\ref{bolla}) together with the regularization of
the contact potential:
\begin{equation}
\bar L({\mathbf q}, \omega) = A({\mathbf q}, \omega) +
v_{\uparrow\downarrow} 
\frac{4 [C({\mathbf q},\omega)]^2}
{1+v_{\uparrow\downarrow}B_{\mbox{\scriptsize reg}}({\mathbf q}, \omega)}\;, 
\label{rpa_final}
\end{equation}
where $A({\mathbf q}, \omega) $, 
$B_{\mbox{\scriptsize reg}}({\mathbf q}, \omega) $ and
$C({\mathbf q},\omega)$ are complex functions of the frequency to be
evaluated 
numerically. We remark that the assumption $\Delta \ll \epsilon_F$ has
considerably simplified the treatment by allowing to introduce only
three basic functions ($A$, $B$ and $C$) in place of six required by
the exact treatment \cite{griffin_rpa}.
 The general expression  $ R({\mathbf q},\omega)$, 
where $R$ stands for $A$, $B$ or $C$  and $r$ stands for $a$, $b$,
$c$,  is given by 
\cite{griffin_rpa}
\begin{eqnarray}
R({\mathbf q}, \omega)= \int \frac{d^3k}{(2 \pi)^3} {r}_L({\mathbf
k},{\mathbf q}) 
(f(E_+)-f(E_-)) \left[\frac{1}{\hbar\omega+ E_+-E_-+ i \eta}- \alpha
\frac{1}{\hbar\omega- (E_+-E_-)+ i \eta}\right] \nonumber \\ +{r}_B({\mathbf
k},{\mathbf q}) (1-f(E_+)-f(E_-))\left[\frac{1}{\hbar\omega- (E_++E_-)+ i
\eta}-\alpha 
\frac{1}{\hbar\omega+E_++E_-+ i \eta}\right]
\label{Gi}
\end{eqnarray}
where the ``Landau'' contributions ${r}_L({\mathbf k},{\mathbf q})$
 are given by 
\begin{eqnarray}
a_L({\mathbf k},{\mathbf q}) &=& (1+(\xi_{+}\xi_{-} -\Delta^2)/E_{+}E_{-})/4 \\
b_L({\mathbf k},{\mathbf q}) &=& (1-(\xi_+\xi_-+\Delta^2)/E_+E_-)/4 \\
c_L({\mathbf k},{\mathbf q}) &=& \Delta(1/E_+-1/E_-)/8
\end{eqnarray}
and where the ``Beliaev'' contributions ${r}_B({\mathbf k},{\mathbf q})$ are 
given by
\begin{eqnarray}
a_B({\mathbf k},{\mathbf q}) &=& (1-(\xi_+\xi_- -\Delta^2)/E_+E_-)/4 \\
b_B({\mathbf k},{\mathbf q}) &=& (1+(\xi_+\xi_-+\Delta^2)/E_+E_-)/4 \\
c_B({\mathbf k},{\mathbf q}) &=& -\Delta(1/E_++1/E_-)/8
\end{eqnarray}
for $A({\mathbf q}, \omega) $, $B({\mathbf q}, \omega) $ and $C({\mathbf q}, \omega) $ respectively. 
The function $B({\mathbf q},\omega)$ as defined in Eq.~(\ref{Gi})
presents an ultraviolet divergence originating
from the choice of a contact interaction potential.
This divergence is removed in a systematic way
by the use of the pseudopotential, which amounts here 
simply to subtracting the most diverging contribution
in the form $1/(2\xi_{\mathbf k})$:
\begin{equation}
B_{\mbox{\scriptsize reg}}({\mathbf q},\omega)=
B({\mathbf q},\omega)
-\int\frac{d^3k}{(2\pi)^3}{\cal P}\left(\frac{1}{2\xi_k}\right)
\end{equation}
where ${\cal P}$ represents the principal value \cite{nota}.
We have used the
notations 
$\xi_\pm=\hbar^2({\mathbf k} \pm {\mathbf q}/2)^2/2m+
v_{\uparrow\downarrow}n/2-\mu$ 
where $n$ is the total equilibrium density of particles,
$E_\pm=(\xi_\pm^2+\Delta^2)^{1/2}$ and 
$f(E)=1/(\exp(\beta E)+1)$ is the Fermi distribution function at
temperature $T$ with 
$\beta= 1/(k_B T)$. 
The parameter $\alpha$  equals 1 for $A({\mathbf q}, 
\omega) $ and $B({\mathbf q}, \omega) $, 
while $\alpha$ equals $-$1 for $C({\mathbf q}, \omega) $, and
 $\eta$ is a positive
infinitesimal.

The final expression for the  double Fourier transform of the
density-density  response function  
in RPA is obtained from
Eq.~(\ref{rpa_proper}) as:
\begin{equation}
\chi({\mathbf q}, \omega)=\frac{2 \bar L({\mathbf q},
\omega)}{1-v_{\uparrow\downarrow}\bar L({\mathbf q}, \omega)} \;.
\label{rpa_uso}
\end{equation}
This can be contrasted with the density-density response function in the static
Bogoliubov approximation, which  leads to 
$\chi({\mathbf q}, \omega)=2 A({\mathbf q},\omega)$ \cite{static_bog}.

\section{The spectrum of density fluctuations}

\subsection{Homogeneous system}

Before displaying the fully 
numerical solution of Eqs.~(\ref{rpa_final}) and (\ref{rpa_uso}), we analyze
some limiting cases.
At temperatures higher than the BCS transition
temperature we have $B({\mathbf q},
\omega)=0$, $C({\mathbf q},\omega)=0 $ and $A({\mathbf q}, \omega)=
\chi_0({\mathbf q}, \omega)$, where $\chi_0({\mathbf q}, \omega)$
is the well-known Lindhard function
for the response of a ideal Fermi gas (see for example \cite{pines_nozieres}).
The RPA equation~(\ref{rpa_final}) reduces to the usual expression
$\chi=2 \chi_0/(1-v_{\uparrow\downarrow}\chi_0)$, the factor two
being due to the two spin components of the gas. 
In the case of
repulsive interactions ({\it i.e.}~$v_{\uparrow\downarrow}>0$) the
equation shows a pole corresponding to the zero sound, while no
well-defined  collective excitation is stable in the case of
attractive interactions.

At zero temperature, in the limit of small ${\mathbf q}$ and $\omega$ it
is possible to estimate analytically the expression for the density
response function; to lowest order in ${\mathbf q}$ we obtain 
\begin{equation}
\chi({\mathbf q},\omega)= \frac{c_B^2q^2{\cal
N}(\epsilon_F)}{(\omega+i \eta)^2-c_{B}^2 q^2(1+2 k_F
a_{\uparrow\downarrow}/\pi)}, 
\label{eq:nouvelle_equation}
\end{equation}
where $c_B=v_F/\sqrt{3}$ is the sound
velocity predicted by Bogoliubov, ${\cal
N}(\epsilon_F)=mk_F/\pi^2\hbar^2$ is the density of states at the Fermi
level and $a_{\uparrow\downarrow}$ is the scattering length
such that $v_{\uparrow\downarrow}=4\pi\hbar^2a_{\uparrow\downarrow}/m$. The pole yields a phonon-like excitation, corresponding to the
Bogoliubov-Anderson  sound for this
system. Eq.~(\ref{eq:nouvelle_equation}) holds approximately until 
the phonon is stable, that is before it meets 
the continuum of quasiparticle-quasihole
excitations, which has a threshold energy of $2 \Delta$.
 Evidently the RPA
expression is valid in the dilute limit $k_F|a_{\uparrow\downarrow}|\ll 1$.

It is easily checked that in the long-wavelenght limit the
Bogoliubov-Anderson  sound
exhausts the $f$-sum rule
$-\int d\omega\,\omega {\rm Im} \chi({\mathbf q},\omega)=\pi n q^2/m$,
where $n$ is the total density of the gas. A more 
general proof can be obtained by noticing that the RPA, being
equivalent to the time-dependent 
Hartree-Fock-Bogoliubov theory, automatically
satisfies the continuity equation and hence the $f$-sum rule.
On the
contrary, the static Bogoliubov approximation results to be
bad in the limit $q\rightarrow 0$: from
Eq.~(\ref{Gi}) we estimate that $A(q\rightarrow
0,\omega)\propto \omega^{-3/2}$ in the high frequency limit, yielding
an infinite contribution to the first moment integral. 

\begin{figure}
\centerline{\psfig{file=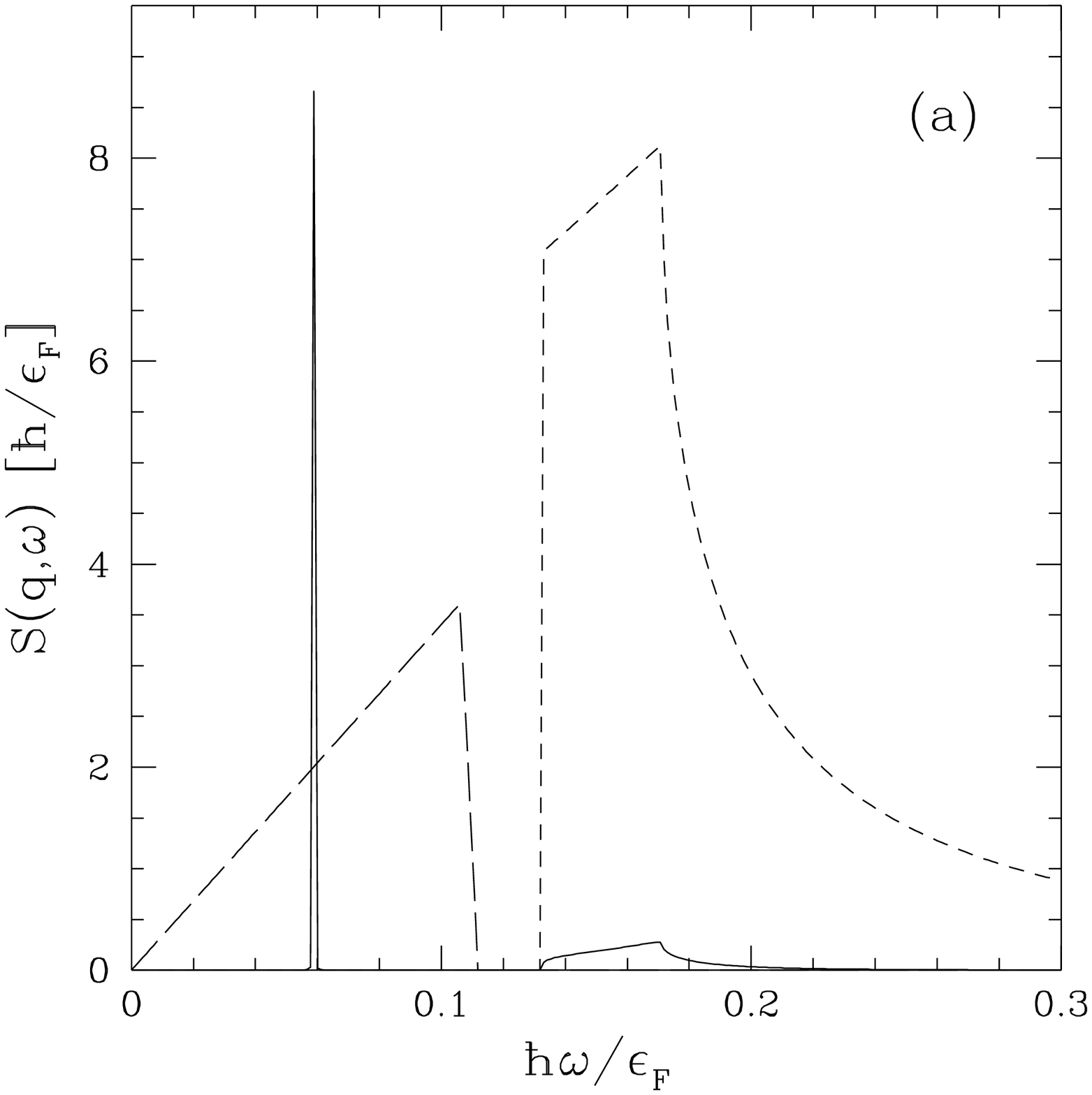,width=0.4\linewidth}\psfig{file=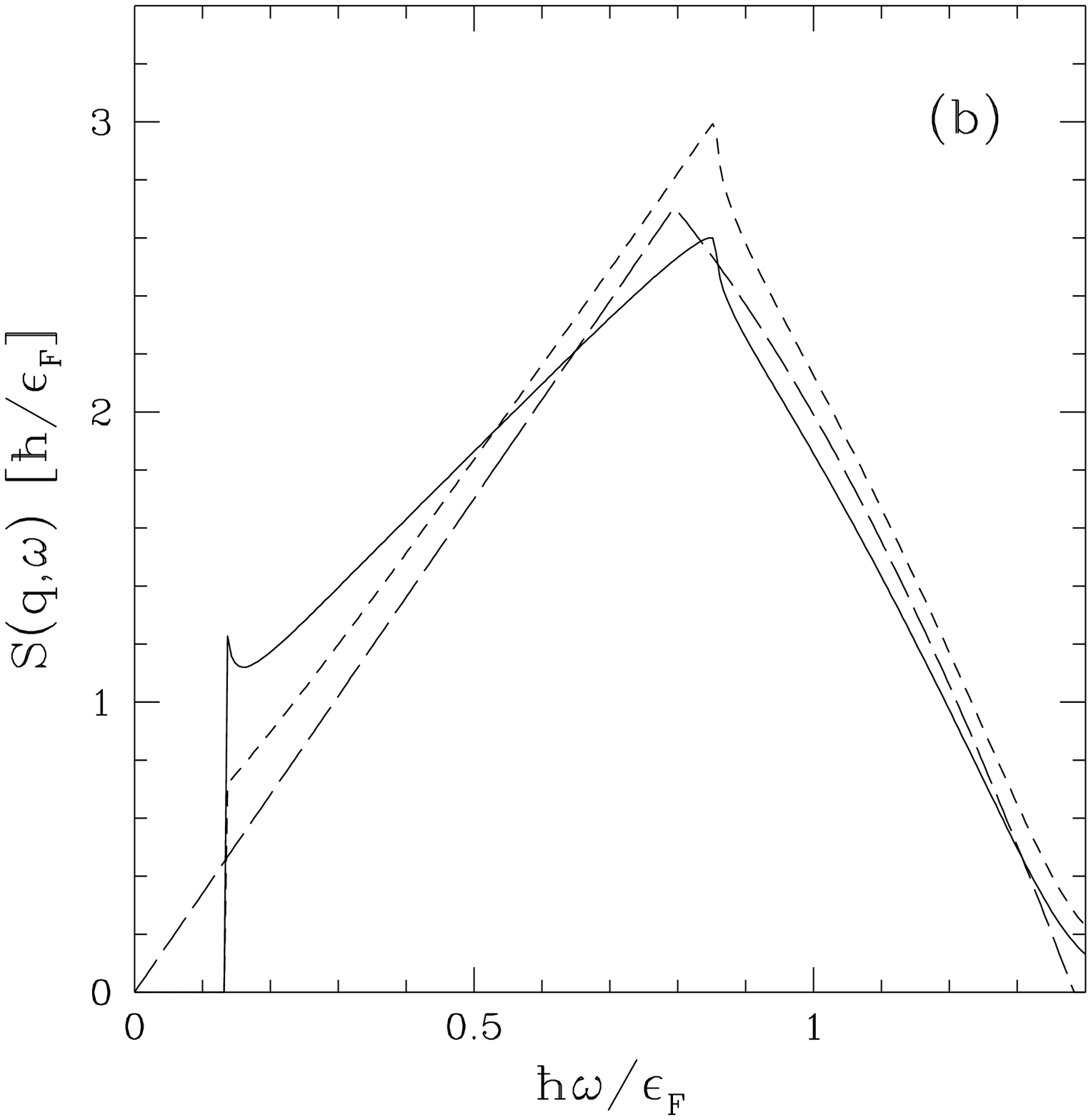,width=0.4\linewidth}}
\caption{Dynamic structure factor of a homogeneous superfluid Fermi
gas at $T=0$ for $q=0.054$ $k_{F}$ (a), and for $q=0.54$  $k_{F}$ (b),
 as predicted by the RPA (solid line)
and by the static Bogoliubov  approximation (dashed line). The results
for a non-interacting Fermi gas (long-dashed line) are also shown. $k_{F}$
is the Fermi wavevector  and 
$\epsilon_{F}=\hbar^2k_F^2/2m$ is the Fermi energy of the
non-interacting Fermi gas at the same density. The
parameters chosen are $\Delta=0.065 \epsilon_{F}$ and $k_F
a_{\uparrow\downarrow}=-0.04$.}  
\label{fig1}
\end{figure}

We now turn to the presentation of numerical results.
Rather than plotting the complex quantity $\chi({\mathbf q},\omega)$ we
have chosen to represent the spectrum of total density fluctuations of 
wavevector ${\mathbf q}$, given by the dynamic structure factor
$S({\mathbf q},\omega)$. On an experimental point of view
the dynamic structure factor can be accessed
{\sl via} the rate of the scattering events of a probe
particle by the gas leading to a momentum exchange $\hbar{\mathbf q}$
and to an energy exchange $\hbar\omega$ between the probe particle
and the gas; on a theoretical point of view the dynamic structure factor
of the gas is related to Im$\chi({\mathbf q},\omega)$ by the
fluctuation-dissipation theorem \cite{pines_nozieres}: 
\begin{equation}
S({\mathbf q},\omega)=-(2\hbar /n) (1-\exp(-\beta \hbar
\omega))^{-1}{\rm Im} \chi({\mathbf q},\omega).
\end{equation}
Figure~\ref{fig1} shows the spectrum of a homogeneous superfluid at
zero temperature as resulting from the full RPA calculation, compared
to the predictions of the static Bogoliubov approximation: it is
evident that for small $q$ ($\hbar q <2 \Delta/c_{B}$) the
Bogoliubov approximation yields an unphysically large response (Fig.~1
(a)). For larger $q$ the Bogoliubov-Anderson phonon falls in the continuum of
quasi-particle quasi-hole excitations, and the two approximations
yield almost 
the same result (Fig.~1 (b)), which is also close to the ideal-gas
solution. 

\subsection{Harmonically trapped system}

We turn now to the situation where the particles are subject to an
external harmonic confinement. We assume that the  confining 
potential has the same action   
on both spin components, this is indeed the case in a laser induced
trap. For simplicity we further assume that the resulting trapping
potential is isotropic so that 
\begin{equation}
V_{\mbox{\scriptsize ext},\uparrow}(r)=
V_{\mbox{\scriptsize ext},\downarrow}(r)= \frac{1}{2} m\Omega^2 r^2.
\end{equation}
We consider bulk excitations of the harmonically confined cloud
induced
by a probe potential $U({\mathbf r},t)$ of wavevector ${\mathbf q}$
and frequency $\omega$. We characterize the density response of the gas
by the dynamic structure factor calculated in
the local-density approximation:
\begin{equation}
S_{LDA}({\mathbf q}, \omega)=\frac{-2 \hbar}{1-e^{-\beta \hbar\omega}}\int d^3r\, {\rm
Im} \chi ({\mathbf q},\omega; \mu(r),\Delta(r)) \;,
\label{slda}
\end{equation}
where $\chi$ is the density response function 
derived in the previous section for the homogeneous
system. This  local-density
approximation  is valid for $q \geq 1/R$, where $R$
is the radius of the cloud, since it does not take into account
surface modes \cite{baranov} and in-gap single-particle  excitations
\cite{baranov2}. The same approach has 
already described successfully an experiment on trapped Bose-Einstein
condensates, where the experiment has directly
measured the function ${\rm Im} \chi ({\mathbf q},\omega)\propto
[S({\mathbf q},\omega)-S(-{\mathbf q},-\omega)]$ \cite{ketterle}.
 
The position-dependent chemical potential and gap in Eq.~(\ref{slda})
are defined as 
$\mu(r)=\mu-V_{\mbox{\scriptsize ext}}(r)$ and $\Delta(r)=\Delta[n(r)]$.  
The chemical  potential  $\mu$ of 
each spin component is determined by the normalization condition  $N=\int
d^3r\, n(r)$ where $N$ is the total number of particles in the gas.
The equilibrium  total density profile $n(r)$ and the gap  $\Delta(r)$ are
obtained first by numerical solution of the BCS equations in the
homogeneous system 
\begin{equation}
n = 2 \int \frac{d^3{k}}{(2 \pi)^3}\left\{ |u_{\bf k}|^2 f(E_{\bf
k}) + |v_{\bf k}|^2 [1 - f(E_{\bf k})] \right\}\;,
\label{n1}
\end{equation}
\begin{equation}
\int \frac{d^3{k}}{(2 \pi)^3}  \left\{ \frac{1- 2 
f(E_{\bf k})}
{2E_{\bf k}} - 
{\cal P} \left(\frac{1}{2\xi_{\bf k}}\right) \right\}
 = - \frac{1}{v_{\uparrow\downarrow}}\;,
\label{gapeq}
\end{equation}       
and then  by employing the Thomas-Fermi approximation (TFA)
\cite{noticina} to take
into account 
the inhomogeneity due to the external confinement. 
  
The main effect of the external confinement is a broadening of  the spectrum, 
which is due to the inhomogeneous distribution of the density in the trap.
This is illustrated already by a simple analytic expression for
the dynamic structure factor at zero temperature in the low $q$ limit:
integrating the imaginary
part of Eq.(\ref{eq:nouvelle_equation}) over space yields
to lowest order in $k_F |a_{\uparrow \downarrow}|$  
\begin{eqnarray}
S_{LDA}^{\mbox{\scriptsize phonon}} ({\mathbf q}, \omega)
=2 \int d^3 r\, A_q[n(r)]
\delta(\omega-\omega_q[n(r)])=  12 \sqrt{3}\,\Omega^{-1}
\left(\frac{\bar \omega}{\bar q}\right)^3 \sqrt{2 \bar
\mu
-3\left(\frac{\bar \omega}{\bar q}\right)^2 }\;.
\label{slda0}
\end{eqnarray}
Here, $A_q[n]=m k_F c_Bq/2\pi\hbar$, $\omega_q=c_Bq$ and we
have adopted the rescaled 
units $\bar
\omega=\omega/\Omega$, $\bar q=\hbar q/\sqrt{m\hbar \Omega}$ and 
$\bar \mu =\mu/\hbar \Omega$.  For the density profile
we have taken  the TFA expression 
$n(r)=(2m(\mu-V_{\mbox{\scriptsize ext}}(r))/\hbar^2)^{3/2}/3\pi^2$, 
which  neglects the Hartree mean field effect 
but turns out to be a good approximation
in the dilute limit \cite{stoof_marianne}. 

\begin{figure}
\centerline{\psfig{file=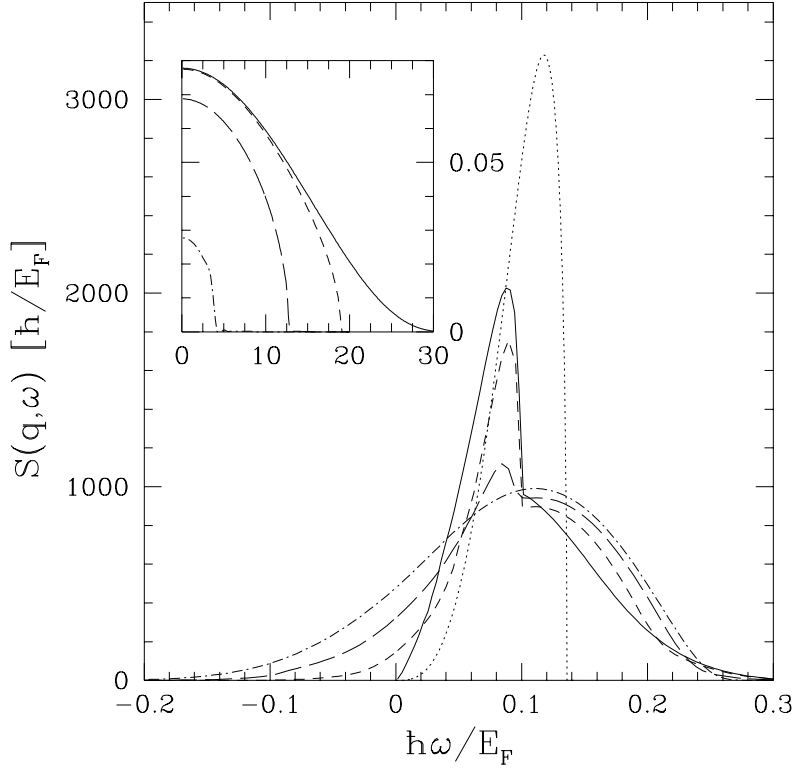,width=0.6\linewidth}}
\caption{Dynamic structure factor of a confined superfluid Fermi gas
for $\hbar q=4.15$ $p_{ho}$
at different temperatures: $T=0$ (solid line), $k_B T=0.014$ $ E_{F}$
(short-dashed line), $k_B T=0.028$ $ E_{F}$ (long-dashed line) and
$k_B T=0.042$ $ E_{F}$ (dot-dashed line). The critical temperature
for the BCS transition is $k_B T=0.047 $ $E_{F}$. The analytic
expression for the broadening of the Bogoliubov-Anderson phonon in the
low $q$  and zero temperature limit,
Eq.~(\ref{slda0}), is
given in dotted line. The inset shows the gap function in units of
$E_F$ as a function of the radial coordinate $r/a_{ho}$, as obtained
from the local-density approximation. The different line-styles
correspond to the  different temperatures of the main figure. 
$p_{ho}=\sqrt{\hbar m \Omega}$, $a_{ho}=\sqrt{\hbar/m\Omega}$ and $E_F$ is the
Fermi energy of a harmonically trapped non-interacting Fermi gas with
the same number of particles. The
parameters used are $N=8 \times 10^7$, 
$\Omega=2\pi \times 27.2$ s$^{-1}$ and $a_{\uparrow\downarrow}=-2160$ $a_0$
where $a_0$ is the Bohr radius.
}
\label{fig2}
\end{figure}

The numerical results at finite temperature and wavevector,
presented in Fig.~\ref{fig2}, show in the
low temperature spectrum  a peak corresponding to the
Bogoliubov-Anderson phonon and include 
a high-frequency tail due to the contribution of multi-particle
excitations. With increasing temperature, the asymmetric feature due
to the Bogoliubov-Anderson 
 phonon becomes less marked and disappears at $T\simeq
\Delta/k_B$, when quasi-particle quasi-hole pairs are
easily excited by thermal fluctuations.

\section{Conclusion}

 In this paper, we have derived for a two-component spin-polarized Fermi
gas a generalized Random-Phase Approximation
 to describe the excitation  spectrum in the
superfluid phase. We have shown that, contrary to the case of bosonic
systems, the predictions of this theory -- valid in the limit $k_F
|a_{\uparrow \downarrow}|\ll 1$ -- are essentially different from
those of the ``static'' Bogoliubov theory, which instead yields an
unphysically large signal at small wavevectors, due to the lack of
local particle conservation. 

The possible experiments that have motivated this theoretical
work are light scattering \cite{ketterle} or scattering of test particles
\cite{gabri,ketterle_test} by a two-component Fermi gas stored
in a dipole trap. The outcome of this type of experiments
is described by the dynamic structure factor.
We have therefore employed the results of the homogeneous system to predict 
in a local density approximation the
dynamic structure factor of a harmonically trapped superfluid at
finite temperature, and we have shown that the Bogoliubov-Anderson
phonon -- main
feature of the superfluid phase in the spectrum of the homogeneous gas
-- would appear also in the response of the trapped system, as an
asymmetric peak. We have investigated the effects of the temperature
on the shape of the response, showing that the Bogoliubov-Anderson
 phonon should remain
visible up to a temperature $T\simeq \Delta/k_B$.
The observation of the Bogoliubov-Anderson phonon in the response of the gas
to a probe beam may therefore
provide a way to detect the 
presence of the superfluid phase in the experiments on alkali Fermi gases.

Our general RPA equations 
are also suitable for a  full description of
the
inhomogeneous system without local density approximation, 
thus allowing  in principle to  take into account the discrete nature
of the eigenmodes  of the trapped gas. This would complete
the static Bogoliubov treatment already performed in \cite{yvan}.

\acknowledgements
A.M. thanks Professor M. P. Tosi for useful discussions and
Professor A. Griffin for drawing her attention to
Ref.~\cite{griffin_rpa}. Supports from Laboratoire Kastler-Brossel and
INFM are acknowledged.

Laboratoire Kastler-Brossel is a unit\'e de recherche de l'Ecole 
normale sup\'erieure et de l'universit\'e Pierre et Marie Curie,
associ\'ee au CNRS.



\begin{thebibliography}{10}
\bibitem{commento} In the dilute limit the superfluid transition
almost does not affect 
the compressibility, because it modifies only slightly the Fermi
sphere \cite{vecchio}.
\end{thebibliography}
\end{document}